\documentclass{article}

\usepackage{PRIMEarxiv}

\usepackage[utf8]{inputenc} 
\usepackage[T1]{fontenc}    
\usepackage{hyperref}       
\usepackage{url}            
\usepackage{booktabs}       
\usepackage{amsfonts}       
\usepackage{nicefrac}       
\usepackage{microtype}      
\usepackage{lipsum}
\usepackage{fancyhdr}       
\usepackage{graphicx}       
\graphicspath{{media/}}     

\usepackage{glossaries}
\usepackage{cleveref}
\usepackage{booktabs}
\usepackage[normalem]{ulem}
\usepackage{verbatim}
\usepackage{pifont}
\usepackage{threeparttablex}
\usepackage{multirow}
\usepackage{subcaption}
\usepackage{caption}
\usepackage{paralist}

\title{SVRF: Efficient Register Storage for Long-Vector Architectures}

\author{
  Francesco Minervini, Lorenzo Deltetto, Osman Unsal, Adrian Cristal \\
  Barcelona Supercomputing Center - BSC-CNS \\
  \texttt{\{francesco.minervini, lorenzo.deltetto, osman.unsal, adrian.cristal\}@bsc.es} \\
}

\newacronym{vpu}{VPU}{Vector Processing Unit}
\newacronym{hpc}{HPC}{High Performance Computing}
\newacronym{svrf}{SVRF}{Scalarized Vector Register File}
\newacronym{dlp}{DLP}{Data Level Parallelism}
\newacronym{isa}{ISA}{Instruction Set Architecture}
\newacronym{isas}{ISAs}{Instruction Set Architectures}
\newacronym{vla}{VLA}{Vector Length Agnostic}
\newacronym{bsc}{BSC}{Barcelona Supercomputing Center}
\newacronym{mvl}{MVL}{Maximum Vector Length}
\newacronym{sram}{SRAM}{Static Random Access Memory}
\newacronym{fsm}{FSM}{Finite State Machine}
\newacronym{dp}{DP}{Double Precision}
\newacronym{rw}{RW}{Read-Write}
\newacronym{war}{WAR}{Write After Read}
\newacronym{waw}{WAW}{Write After Write}
\newacronym{frl}{FRL}{Free Register List}
\newacronym{fsrl}{FSRL}{Free Scalar Register List}
\newacronym{fml}{FML}{Free Mask List}
\newacronym{rat}{RAT}{Register Alias Table}
\newacronym{rmt}{RMT}{Register Mapping Table}
\newacronym{vcu}{VCU}{Vector Control Unit}
\newacronym{lmu}{LMU}{Load Management Unit}
\newacronym{smu}{SMU}{Store Management Unit}
\newacronym{imu}{IMU}{Item Management Unit}
\newacronym{vrf}{VRF}{Vector Register File}
\newacronym{fpu}{FPU}{Floating Point Unit}
\newacronym{fpus}{FPUs}{Floating Point Units}
\newacronym{alu}{ALU}{Arithmetic Logic Unit}
\newacronym{simd}{SIMD}{Single Instruction Multiple Data}
\newacronym{vrob}{VROB}{Vector ReOrder Buffer}
\newacronym{fpga}{FPGA}{Field Programmable Gate Array}
\newacronym{rtl}{RTL}{Register Transfer Level}
\newacronym{hdl}{HDL}{Hardware Description Language}
\newacronym{ovi}{OVI}{Open Vector Interface}
\newacronym{uvm}{UVM}{Universal Verification Methodology}
\newacronym{gls}{GLS}{Gate-Level Simulation}
\newacronym{sdf}{SDF}{Standard Delay Format}
\newacronym{bb}{BB}{Body Biasing}
\newacronym{pnr}{PnR}{Place and Route}
\newacronym{8t}{8T}{8-Tracks}
\newacronym{12t}{12T}{12-Tracks}
\newacronym{blas}{BLAS}{Basic Linear Algebra Subprograms}
\newacronym{mmul}{MMUL}{Matrix-Matrix Multiplication}
\newacronym{fft}{FFT}{Fast Fourier transform}
\newacronym{spmv}{SPMV}{Sparse Matrix-Vector Multiplication}
\newacronym{spu}{SPU}{Scalar Processing Unit}
\newacronym{fma}{FMA}{Fused Multiply-Add}
\newacronym{rvv}{RVV}{RISC-V Vector Extension}
\newacronym{hpl}{HPL}{High Performance Linpack}
\newacronym{dsp}{DSP}{Digital Signal Processing}
\newacronym{ml}{ML}{Machine Learning}
\newacronym{raw}{RaW}{Read after Write}
\newacronym{csr}{CSR}{Control and Status Register}
\newacronym{ve}{VE}{Vector Engine}
\newacronym{sve}{SVE}{Scalable Vector Extension}
\newacronym{avx}{AVX}{Advanced Vector Extension}
\newacronym{vlen}{VLEN}{number of bits in a single vector register}
\newacronym{lmul}{LMUL}{vector register group multiplier}
\newacronym{elen}{ELEN}{maximum size in bits of a vector element}
\newacronym{sew}{SEW}{Selected Element Width}
\newacronym{flop}{FLOP}{floating-point operations}
\newacronym{fifo}{FIFO}{First-In First-Out}
\newacronym{dut}{DUT}{Device-Under-Test}
\newacronym{slvt}{SLVT}{Super Low $V_t$}
\newacronym{lvt}{LVT}{Low $V_t$}
\newacronym{pdk}{PDK}{Process Design Kit}
\newacronym{suf}{SUF}{Scalar Utilization Factor}

\glsdisablehyper

\begin{document}
\maketitle

\begin{abstract}
Vector processors exploit data-level parallelism to provide high computational throughput, but very long vectors can make the \gls{vrf} a significant source of area, energy, and register pressure. This cost is exacerbated when vector instructions operate on values with scalar semantics: although such values are architecturally represented using vector registers, they do not require full-width storage.

This paper proposes the \gls{svrf}, a microarchitectural enhancement that redirects scalar-semantics vector values to a dedicated scalar register file, avoiding unnecessary allocation and access of full-width physical vector registers. The \gls{svrf} integrates with vector register renaming while preserving compatibility with the supported vector ISA.

We implement and evaluate \gls{svrf} in an \gls{rtl}-based \gls{rvv}-compliant vector processing unit supporting very long vectors. Across representative HPC and AI/ML workloads, the \gls{svrf} reduces vector-register pressure and \gls{vrf} activity without degrading performance. By exploiting the resulting reduction in \gls{vrf} demand, the proposed design achieves up to 15.8\% \gls{vrf} area reduction and up to 6\% total area reduction, while reducing total power by up to 12\% in evaluated configurations. Some workloads also achieve up to 1.12$\times$ speedup.
\end{abstract}

\keywords{RVV \and Vector Processing \and HPC}

\section{Introduction}
As vector lengths scale to meet the throughput demands of modern scientific and machine learning workloads, the \gls{vrf} is emerging as a primary scalability bottleneck in vector processing units \cite{Kozyrakis2003}. In long-vector implementations, prevalent in \gls{hpc} vector processor designs \cite{nec,vp2000,CRAY-1,Cavalcante2020,Minervini2023}, individual physical vector registers may span several kilobytes, making register storage capacity, access energy, and renaming pressure dominant contributors to area and power consumption. In such designs, improving \gls{vrf} efficiency is central to sustaining energy-efficient performance.


While prior work has focused on reducing \gls{vrf} access energy and improving register banking or port efficiency \cite{Gobieski2019,Titopoulos2025,Ramirez2020}, a more fundamental inefficiency remains underexplored. A substantial fraction of performance-critical vector instructions exhibit inherently scalar semantics. Operations such as scalar-to-vector moves, broadcast gathers, and vector reductions either replicate a single scalar value across whole vector registers or produce a scalar result. These instructions are pervasive in both \gls{hpc} kernels \cite{Zhong2022} (e.g., dot products, norm computations, stencil updates) and AI/ML workloads \cite{Absar2025} (e.g., GEMM accumulation, normalization, attention reductions).

Despite their scalar nature, all modern vector \gls{isas}, such as the Intel \gls{avx}\cite{avx512}, the ARM \gls{sve}\cite{Stephens2017}, and the \gls{rvv}\cite{rvv_specs} define these operations to allocate and access full-width vector registers. As the hardware vector length increases, this semantic mismatch leads to disproportionate \gls{vrf} occupancy, increased physical register pressure, and unnecessary dynamic energy consumption. 

This paper argues that scalar-semantics vector instructions represent a structural inefficiency in long-vector architectures. We observe that not all vector operations require full-width storage and that aligning storage allocation with instruction semantics can significantly reduce \gls{vrf} pressure. To exploit this opportunity, we introduce the \gls{svrf}, a dedicated scalar storage structure integrated within the \gls{vpu}. The \gls{svrf} stores scalar values associated with broadcast and reduction operations and distributes them to the vector lanes on demand, avoiding unnecessary full-width \gls{vrf} allocation while remaining fully compliant with the ISA.


The proposed mechanism integrates seamlessly with vector register renaming through lightweight extensions to allocation and mapping structures and incurs minimal hardware overhead. We evaluate our approach on a \gls{rvv}-1.0-compliant \gls{vpu} supporting vector registers of up to 16384 bits. Across representative \gls{hpc} and AI/ML workloads, the \gls{svrf} reduces \gls{vrf} accesses and physical register pressure, enabling up to 1.12X performance improvement and 12\% total power reduction with minimal area overhead. Furthermore, the \gls{svrf} enables reducing the number of physical vector registers without performance degradation, achieving up to 15.8\% of \gls{vrf} area reduction and improving the scalability of long-vector designs.

This paper makes the following contributions:

\begin{itemize}
    \item We identify scalar-semantics vector instructions as a structural scalability and energy bottleneck in long-vector implementations.
    \item We propose the \gls{svrf}, a microarchitectural enhancement that decouples scalar-oriented vector operations from the full-width \gls{vrf} while preserving ISA compliance. To the best of our knowledge, this is the first research work that explores the idea of a scalar register file embedded in a \gls{vpu}.
    \item We describe its integration with vector register renaming in a \gls{rvv}-compliant \gls{vpu} using lightweight modifications to allocation and mapping structures.
    \item We implement and integrate the mechanism in the \gls{rtl} design of the \gls{vpu}. We evaluate the impact of the integrated modules through all stages of the design up to the physical implementation.
    \item We demonstrate improved area and energy efficiency across representative HPC and AI/ML workloads without affecting execution performance. The results are reported based on the actual implementation.
    \item We posit that our mechanism can be implemented in any vector implementation equipped with register renaming, regardless of the supported vector ISA.
\end{itemize}

The remainder of the paper is organized as follows. \Cref{background} provides background on vector architectures and \gls{vrf} design considerations. \Cref{architecture} describes the proposed \gls{svrf} microarchitecture and its integration with the \gls{vpu} pipeline. \Cref{methodology} presents the experimental methodology for the evaluation of our work. \Cref{evaluation} reports the results of the evaluation and discusses design trade-offs. \Cref{sota} analyzes related work and finally \Cref{conclusion} concludes this paper.


\section{Background} \label{background}

\subsection{Evolution of Vector Architectures}
Vector architectures have historically played a central role in \gls{hpc}. Rather than issuing independent scalar operations, vector machines apply a single instruction to long streams of data elements stored in dedicated vector registers. Early systems such as the CDC STAR-100 \cite{CDC-STAR100} and later the Cray-1 \cite{CRAY-1} demonstrated that this model could sustain significantly higher arithmetic throughput than scalar processors by deeply pipelining functional units and amortizing instruction overhead across many data elements. A key architectural step was the introduction of explicit vector register files, which enabled efficient functional unit utilization and reduced pressure on memory bandwidth. Combined with techniques such as memory interleaving and vector chaining, these designs established vector processing as a dominant paradigm in supercomputing throughout the 1970s and 1980s \cite{Espasa1998}.

Although advances in CMOS technology and the rise of superscalar microprocessors shifted industry focus toward general-purpose out-of-order cores, the underlying principle of exploiting structured data-level parallelism never disappeared~\cite{Asanovic1996,AsanovicPhd}. In recent years, the rapid growth of both HPC simulations and AI/ML workloads—many of which operate on large, regular data structures—has renewed interest in vector processing as a scalable and energy-conscious architectural approach \cite{Ramirez2020}.

\subsection{Scalable Vector Architectures}\label{bg:varch}

Modern vector \gls{isas} extend classical vector principles while explicitly addressing portability and scalability across implementations. In contrast to fixed-width SIMD extensions, such as the Intel \gls{avx}-512 \cite{avx512}, scalable vector \gls{isas} decouple the architectural vector length from its physical hardware implementation. 
Examples of such architectures include the NEC vector extension \cite{nec}, the ARM \gls{sve} \cite{Stephens2017}, and the \gls{rvv} \cite{rvv_specs}. The \gls{rvv} architecture has recently reached a major milestone with the ratification of version 1.0 of the ISA. Its open, royalty-free specification has significantly lowered barriers to entry, accelerating both academic research \cite{Perotti2024,Minervini2023,Purayil2025} and industrial adoption \cite{smd_vpu,sifive}. One of the key features of the \gls{rvv} is that it is a \gls{vla} ISA. This means that a program written to run on a certain \gls{rvv}-compliant vector processor can still run without modifications on a different \gls{rvv}-compliant vector processor. As such, \gls{rvv} exposes the vector length, VLEN, an implementation-dependent parameter, allowing hardware designers to support relatively short vectors or extend to very long vectors, up to a maximum of 65536 bits per vector register.

A substantial subset of operations common to scalable vector \gls{isas} exhibits scalar semantics within the vector execution model. Representative examples include scalar-to-vector broadcasts, gather instructions indexed by scalar values, and reduction operations that accumulate vector elements into a scalar result. 
These operations are fundamental across application domains. In \gls{hpc} workloads, reductions are central to dot products \cite{dot_prod}, norm computations\cite{norm_comp}, and iterative stencil updates \cite{stencil_vector}. In AI/ML applications, matrix multiplication kernels rely on scalar accumulations \cite{ai_gemm}, while normalization and attention mechanisms frequently combine reductions with subsequent broadcasts\cite{ai_norm}. In conventional implementations, the results of these operations are stored in full-width vector registers, even when the data being represented corresponds to a single scalar replicated across lanes or produced as a scalar outcome. This design choice implies that the storage cost of representing scalar quantities scales proportionally with the vector length. The implications of this scaling become more pronounced as vector widths reach multiple kilobytes per register.

\subsection{Vector Register File Design Challenges}

The \gls{vrf} is a central structure in vector processors. Its organization and scaling characteristics have a direct impact on performance, area, and energy efficiency \cite{Titopoulos2025}. 

As the  vector length increases, each architectural vector register maps to a correspondingly wider physical storage structure. Consequently, even a moderate number of physical vector registers can represent a substantial fraction of the total on-chip memory \cite{Cavalcante2022}, dramatically affecting both dynamic and static power consumption \cite{Mavroidis2001,Kozyrakis2003}.
Beyond raw capacity, bandwidth requirements further complicate \gls{vrf} design. To sustain parallel execution across multiple vector lanes, the \gls{vrf} must support multiple concurrent read and write accesses. Achieving this through multiported structures increases routing complexity and area, and may introduce timing challenges \cite{Kozyrakis2003}. Banking strategies can mitigate port pressure, but introduce their own trade-offs in terms of conflict management and layout complexity \cite{Purayil2025}.

In designs that incorporate register renaming \cite{Espasa1997}, the \gls{vrf} must provision a sufficient number of physical vector registers to expose instruction-level parallelism and avoid \gls{waw} and \gls{war} hazards. Increasing the physical register count, however, directly inflates storage capacity and exacerbates energy consumption. For instance, the NEC VE20 architecture \cite{nec_bis} supports register renaming by including 256 physical vector registers, each composed of 256 \gls{dp}-elements. Therefore, the \gls{vrf} accounts for 512 kB of storage in a single vector core, comparable to the sizes of private caches of modern out of order processors.

Prior research has explored several approaches to reducing \gls{vrf} overhead, including hierarchical organizations \cite{Ramirez2020}, banked designs \cite{Mavroidis2001,Razilov2025} and port reduction techniques \cite{Gobieski2019}. These techniques primarily focus on optimizing access patterns or reducing per-access energy cost. However, they generally assume that all vector results require full-width storage allocation. As vector lengths grow, mechanisms that reduce not only access cost but also unnecessary storage allocation become increasingly relevant for sustaining scalable and energy-efficient vector processing.
\section{Scalar-Aware Vector Register Architecture}\label{architecture}

\Cref{bg:varch} highlighted that scalar-semantics vector instructions incur storage costs that scale with the vector length despite having constant logical data footprint. To address this inefficiency, this Section presents a scalar-aware microarchitectural extension that decouples scalar-valued vector operands from the full-width \gls{vrf}. The proposed design introduces a lightweight scalar storage structure, named \gls{svrf}, dedicated to holding the scalar results. By allowing these values to bypass full-width vector register allocation, the architecture reduces \gls{vrf} pressure without altering the architectural vector execution model. 

The remainder of this Section details the design of the proposed scalar-aware register architecture. We first outline the architecture of the baseline \gls{vpu} where the microarchitectural extension is integrated. We then introduce the design concepts of the proposed architecture. In particular, we describe the extensions to the register renaming stage of the baseline \gls{vpu} design to track the \textit{scalarized} vector results.. We then present the scalar storage structure and its organization, discussing its integration in the vector pipeline and instruction execution.

\subsection{Baseline Vector Microarchitecture}\label{vitruvius}
The proposed scalar-aware register architecture is incorporated into a baseline architecture which features a decoupled \gls{rvv}-1.0-compliant \gls{vpu} design, similar to previous implementations \cite{Espasa1996}\cite{Perotti2024}\cite{Purayil2025}\cite{Minervini2023}. The baseline architecture supports out of order execution through register renaming at vector-register granularity and scalable long-vector configurations.

\begin{figure*}[htb]
    \centering
    \includegraphics[width=\textwidth]{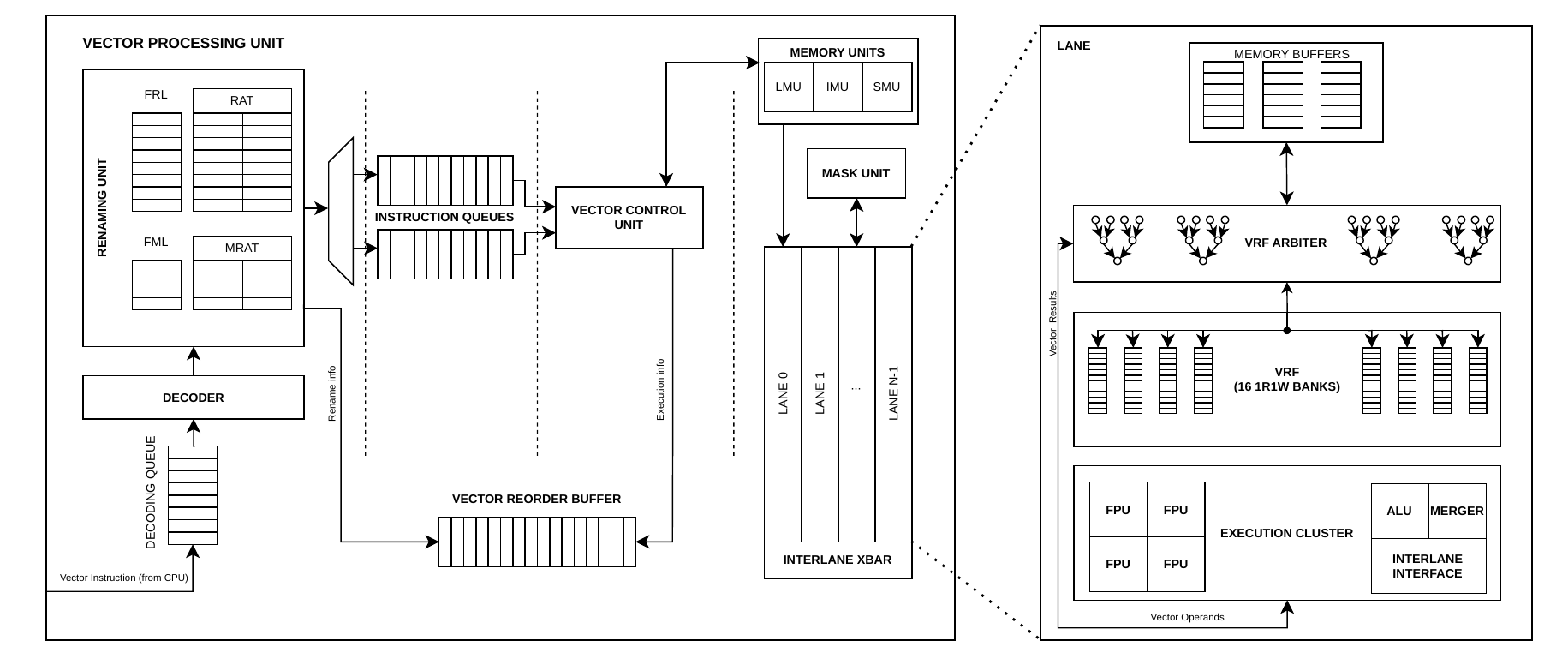}
    \caption{Block diagram of the baseline VPU architecture. The renaming stage is composed of the Free Register List (FRL); the Free Mask List (FML); the Register Alias Table (RAT); the Mask Register Alias Table (MRAT). The memory units are the Load Management Unit (LMU); Index Mask Unit (IMU); Store Management Unit (SMU). The lanes include a Vector Register File composed of 16 two-port memories and related arbitration unit; four Floating-Point Units (FPU); one Arithmetic Logic Unit (ALU); a Merger unit; an interface to the interlane crossbar.}
    \label{fig:baseline}
\end{figure*}

The \gls{vpu} natively targets long-vector execution, with each architectural vector register holding up to 2 kB of data (VLEN=16384 bits), with the maximum size of a vector element ELEN=64 bits. The default configuration instantiates eight vector lanes to scale datapath bandwidth proportionally to vector width. Each lane integrates four \gls{fpus}, one integer/fixed-point \gls{alu}, and a merger unit to support \gls{rvv} predication and element-wise masking operations. The vector lanes exchange data through a single-cycle latency crossbar interlane interconnect. The \gls{vpu} connects the memory subsystem of the attached scalar core through dedicated 512-bit interfaces, for both vector load and store operations, similar to the \gls{ovi} standard \cite{ovi}.

To sustain the aggregate bandwidth demand of the functional units, the \gls{vrf} is distributed across lanes and implemented as 16 two-port memory banks. Each bank stores 128 entries of 64-bit elements, and a dedicated arbiter orchestrates the accesses to/from the functional units. \Cref{fig:baseline} shows the architecture of the baseline \gls{vpu}, with an overview of the internal structures of the vector lane.

The register renaming subsystem maintains a pool of up to 64 physical vector registers. In the baseline design, all vector instructions—including reductions, scalar broadcasts, and scalar-indexed register gathers—allocate full-width physical vector registers, irrespective of their scalar semantics. The \gls{frl} maintains the available physical vector registers that can be assigned as new physical destinations of the vector instructions. It is implemented as a 64-entry \gls{fifo} queue, allowing the renaming of one instruction per cycle.
The \gls{rat} maps each architectural vector register to the assigned physical register. Whenever an instruction enters the renaming stage, it accesses the \gls{rat} to get the logical-to-physical register translation before proceeding to the issue stage. The \gls{rat} is replicated as many times as the number of entries in the \gls{vrob}. This enables a fast recover of the architectural state of the \gls{vrf} in case of a rollback event, either caused by a trap or an exception.

\subsection{Design Overview}\label{arch:design}

The fundamental goal of the proposed architecture is to decouple scalar-semantics vector values from full-width vector register storage without altering the architectural execution model. Rather than allocating a complete physical vector register for results produced by scalar-semantics instructions, the design introduces a dedicated scalar register file structure within the \gls{vpu}. This additional structure is named \gls{svrf} and is completely oblivious to the software, which continues to manipulate architectural vector registers.

At a high level, the scalar-semantics instructions are categorized in two types:
\begin{itemize}
    \item Rename-resolved instructions include all sorts of scalar-to-vector broadcast instructions and are completely resolved in the renaming stage of the \gls{vpu}. These instructions include: scalar-to-vector move instructions, specifically the variants \texttt{vmv.v.x}, \texttt{vfmv.v.f} and \texttt{vmv.v.i}, which move either a scalar or an immediate value in a whole vector register; and the scalar-to-element-zero vector move instructions, which move an integer or a floating-point scalar value in the element 0 of a vector register, \texttt{vmv.s.x} and \texttt{vmv.s.f}, respectively.
    \item Execution-resolved instructions. This class of instructions include all vector reduction instruction variants, as well as scalar-indexed vector register gather instructions,\linebreak \texttt{vrgather.vx} and \texttt{vrgather.vi}, where the index is either a scalar or an immediate value, respectively. These instructions rely on the execution pipelines inside the vector lanes to produce the scalar-valued vector results
\end{itemize}

An alternative design point would reuse the scalar core register file to store scalar-semantics vector results. However, such an approach  would 
\begin{inparaenum}
    \item increase pressure on the scalar renaming subsystem, \item require additional metadata to track vector-specific execution context, and \item introduce communication overhead between the scalar core and the \gls{vpu}.
\end{inparaenum}
Instead, we introduce the \gls{svrf} within the \gls{vpu} to preserve renaming independence and maintain execution locality.

\begin{figure}[htb!]
    \centering
    \includegraphics[width=\linewidth]{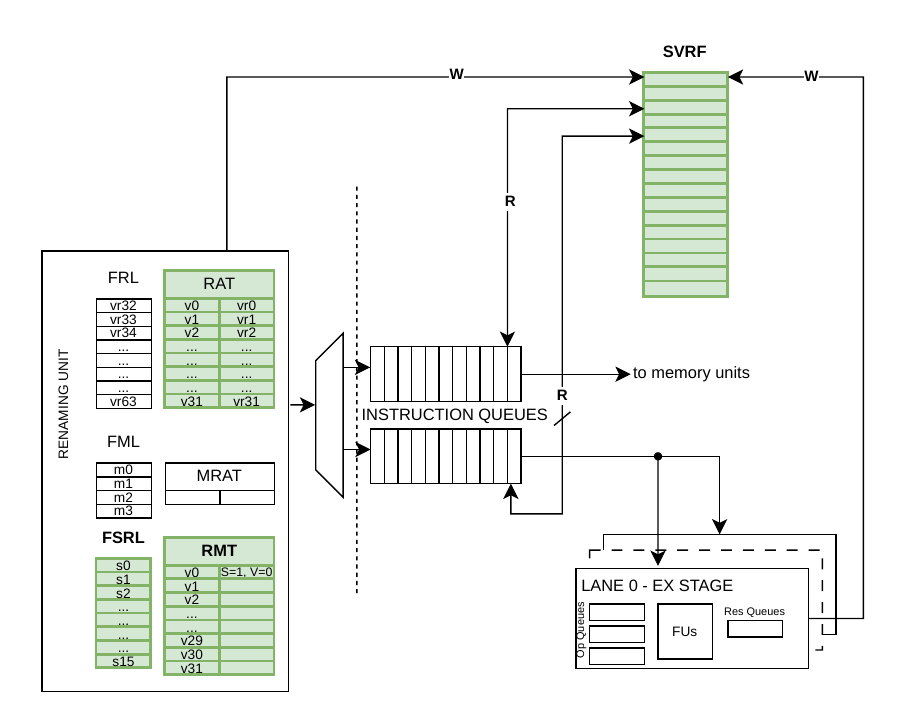}
    \caption{Integration of the SVRF structures in the VPU pipeline.}
    \label{fig:svrf}
\end{figure}

\Cref{fig:svrf} illustrates the high-level integration of the scalar storage structures within the vector pipeline. The next Sections detail the organization of the \gls{svrf}, its interaction with register renaming, and the execution flow modifications required to support scalar-aware allocation.

\subsection{Register Renaming Extensions }\label{arch:renaming}

The register renaming stage of the baseline microarchitecture described in \Cref{vitruvius} is equipped with additional structures to manage the mapping of the vector register to the \gls{svrf}.

The \gls{fsrl} is the scalar equivalent structure of the \gls{frl}. It tracks the availability of scalar registers of the \gls{svrf} instead of physical registers in the \gls{vrf}. Similar to the \gls{frl}, it is implemented as a simple \gls{fifo} queue with as many entries as the number of physical scalar registers in the \gls{svrf}.

When a scalar-semantics vector instruction reaches the renaming stage, the rename control logic checks for the availability of scalar registers in the \gls{fsrl}. If the \gls{fsrl} is not empty, the vector destination register is assigned to a physical scalar register entry in the \gls{svrf}. In the eventuality that the \gls{fsrl} does not have any available physical scalar register, the scalar-semantics instructions are allowed to rename the destination vector register to one of the available physical vector registers from the \gls{frl} as other vector instructions. This ensures that the front-end pipeline is not stalled and incoming instructions can proceed to the execution stage of the \gls{vpu}.

Whenever a new scalar physical register is assigned, the logical-to-physical vector register mapping is registered in the corresponding entry of the \gls{rat}, as described in \Cref{vitruvius}. In order to distinguish whether the vector registers are mapped to a scalar or vector physical register, the register renaming subsystem is added a dedicated structure called \gls{rmt}. This table includes as many 1-bit entries as the number of architectural vector registers. When the bit in a specific entry is set to 1, the corresponding vector register is mapped to a physical scalar register in the \gls{svrf}. Otherwise, it maps to a physical vector register in the \gls{vrf}.

\subsection{Scalarized Vector Register File }\label{arch:svrf}

Scalar values produced as destinations of scalar-semantics instructions are stored in the \gls{svrf}, a microarchitectural-only visibile register file.

The \gls{svrf} is implemented as a small multi-ported register file. Each entry is 64 bits wide, allowing it to store one scalar element at maximum supported precision (ELEN = 64). Narrower elements occupy the least significant bits of the entry. The total number of entries is defined via a specific configuration parameter of the \gls{rtl}.

As depicted in \Cref{fig:svrf}, the \gls{svrf} is placed near the issue stage of the vector pipeline to enable low-latency access by both the renaming logic and the execution lanes. As described in \Cref{arch:design}, it is provisioned with two write ports to be able to concurrently receive results from both the register renaming stage and the execution stage of the \gls{vpu}. In addition, it is equipped with five read ports, which correspond to the maximum number of scalar-consuming accesses that can be issued per cycle. This is because the read accesses occur at the issue stage, from both the arithmetic and the memory instruction queues:
\begin{itemize}
    \item Three operands for the most demanding arithmetic instructions, such as multiply-accumulate instructions.
    \item Two operands for the most demanding memory instructions, such as vector scatter-indexed store-instructions.
\end{itemize}
The \gls{svrf} associates a valid bit for each register. The valid bit is set when a new scalar result is written to the corresponding entry. When a new read request for an entry reaches the \gls{svrf}, the valid bit is polled. The scalar value is provided as response in the next cycle if the valid bit is set, allowing the requesting instruction to proceed to the execution stage inside the lanes.

\subsection{Scalar-semantics Instruction Execution}\label{arch:exec}
Scalar-semantics instructions follow distinct execution flows depending on whether their results are resolved at rename time or execution time. This Section describes how scalarized vector registers are consumed and produced within the execution stage of the vector pipeline. 

Rename-resolved instructions generate values that are fully determined at the renaming stage, as described in \Cref{arch:design}. For these instructions, the scalar value is written directly into the selected entry of the \gls{svrf}, and no execution within the vector lanes is required to produce the result. As such, these instructions are marked as completed in the \gls{vrob} and removed from the pipeline, considerably reducing execution time.

Execution-resolved instructions rely on the vector lanes to compute their scalar result. These instructions execute identically to the baseline design. Upon completion, the final scalar result is delivered to the lane 0, which always keeps the first element of each vector register. However, instead of writing the result into the \gls{vrf}, the writeback stage redirects the scalar outcome to the assigned entry in the \gls{svrf}. The reduction datapath and lane-local accumulation logic, as well as the communication with the interlane interconnect, is identical to the baseline microarchitecture.

When an instruction reads a source operand mapped to the \gls{svrf}, the operand fetch logic selects the scalar read path instead of accessing the \gls{vrf} banks. The scalar value is retrieved during issue and distributed to the vector lanes, together with the per-source metadata to replicate the values accordingly. Consequently, no additional \gls{vrf} bandwidth is consumed for scalarized operands.


Importantly, the execution pipelines are unaware of whether an operand originates from the \gls{svrf} or the original \gls{vrf}. From the perspective of the vector lane control logic, scalarized operands are indistinguishable from immediate operands or scalar-core forwarded values.
\section{Methodology} \label{methodology}
This Section presents the experimental setup for the evaluation of our work. We first describe the \gls{rtl} implementation of the register file extension and the integration to the rest of the \gls{vpu}. Then, we introduce the experimental platforms for the functional validation of our proposal. We also present the evaluation benchmarks and the metrics for the overall assessment of our work.

\subsection{RTL Implementation}\label{eval:rtl}

We use the System Verilog \gls{hdl} for the \gls{rtl} implementation of the \gls{svrf} and related mechanisms in the register renaming and execution stages of the \gls{vpu}. 

The array of registers in the \gls{svrf} is based on a fully configurable \gls{rtl} model of a flip-flop based memory array. This model exposes architectural characteristics as instantiation parameters. Therefore, we can rapidly evaluate different configurations of the \gls{svrf} by simply modifying the parameters in the module instantiation.

The modifications in the register renaming and execution stages integrate seamlessly with the pipeline of the baseline \gls{vpu}. The additional structures in the register renaming unit do not add latency to the front-end, maintaining the renaming throughput of the original implementation. The execution stage requires the addition of the control signals from the instruction queues to the \gls{svrf} in order to retrieve the scalar values associated to previously scalarized vector source operands. Reading the operands from the \gls{svrf} requires one clock cycle. However, this happens in parallel with the validation of the other operands of the head instruction in the queue, therefore it does not introduce additional latency on the handshakes with the vector lanes.

\subsection{Experimental Setup}\label{eval:setup}
In this Section, we describe the \gls{uvm}-based environments for the validation of our work. We use these platforms for the evaluation metrics presented in \Cref{eval:metrics}. We use Siemens Questa Sim \textregistered 2024.3 to run our experiments.

\subsubsection{Standalone VPU Simulation}\label{eval:setup:vpu}
The \gls{rtl} implementation described in \Cref{eval:rtl} is functionally validated in the \gls{uvm} verification infrastructure of the \gls{vpu}. In this environment, the \gls{vpu} represents the \gls{dut} and the \gls{uvm} components generate sequences of stimuli to stress the whole architecture. This platform is crucial for identifying the bottlenecks of the \gls{vpu} architecture exclusively, since it idealizes all the rest of the system. We stress the integrated architecture both through a set of synthetic tests representative of the \gls{rvv} specifications, as well as with pseudo-randomly generated tests with more than 30000 vector instructions.
\subsubsection{System-Level Simulation}\label{eval:setup:system}
The \gls{vpu} is integrated with an in-house out-of-order scalar core in a system which includes the L1 instruction and data caches. The interface between the \gls{vpu} and the scalar core is built on top of the original \gls{ovi} specifications \cite{ovi}. The scalar core is treated as an unmodified black-box component and is not the focus of this work.
The integrated system is validated in its specific \gls{uvm} environment which includes a model of main memory. 

\subsection{Evaluated Configurations.}\label{eval:configs}
The configurable \gls{rtl} implementation of the \gls{svrf} discussed in \Cref{eval:rtl} allows us for an extensive evaluation of the idea over three dimensions:
\begin{itemize}
    \item number of physical vector registers in the \gls{vrf}.
    \item number of physical scalar registers in the \gls{svrf}.
    \item the size of each physical vector register, VLEN.
\end{itemize}
In our experiments, we utilize the following state space exploration process: we vary one of the microarchitectural elements above while keeping the others fixed; then, we combine them to extensively characterize our mechanism for the target workloads. \Cref{table:configs} reports the configurations analyzed in this work. The configurations with the prefix VLEN16k-V64 refer to either the original architecture of the \gls{vpu} with VLEN=16384 bits described in \Cref{vitruvius}, or variants of it with the inclusion of different configurations of the \gls{svrf}. In these cases, the configuration is identified by the suffix including the letter "S" followed by a number indicating the amount of physical scalar registers in the \gls{svrf}. For example, when the \gls{svrf} is configured with 8 physical scalar registers on top of the baseline architecture, the corresponding configuration is VLEN16k-V64-S8. We also study if the introduction of the physical scalar registers in the \gls{svrf} allows us to reduce the number of physical vector registers in the \gls{vrf} without substantial performance degradation. Furthermore, we analyze the impact of the \gls{svrf} for smaller values of VLEN. For this design space exploration, we proceed as follows: first, we reconfigure the \gls{vpu} architecture by scaling down VLEN while maintaining the total computational throughput and the aggregate bandwidth between the \gls{vrf} and the functional units. Since reducing VLEN decreases the amount of simultaneously live vector state required to sustain throughput, we then determine, for each VLEN configuration, the minimum number of physical vector registers that preserves performance. This methodology results in configurations with 56 and 48 physical vector registers for VLEN=8192 and VLEN=4096, respectively. Finally, we evaluate the impact of varying the number of physical scalar registers in the \gls{svrf} on both performance and area. The evaluated configurations are summarized in Table I and are identified by the prefixes  VLEN8k and VLEN4k, respectively.  

\begin{table}[htb]
  \centering
  \caption{Evaluated configurations.}
  \label{table:configs}
  \begin{tabular}{lccc}
    \hline
    \textbf{Config} & \textbf{VLEN} & \textbf{VREGs} & \textbf{SREGs} \\
    \hline
    VLEN16k-V64 & 16384 & 64 & 0 \\
    VLEN16k-V64-S8 & 16384 & 64 & 8 \\
    VLEN16k-V64-S16 & 16384 & 64 & 16 \\
    VLEN16k-V64-S32 & 16384 & 64 & 32 \\
    VLEN16k-V48-S16 & 16384 & 48 & 16 \\
    VLEN16k-V48-S8 & 16384 & 48 & 8 \\
    VLEN16k-V40-S16 & 16384 & 40 & 8 \\
    VLEN8k-V56 & 8192 & 56 & 0 \\
    VLEN8k-V48-S8 & 8192 & 48 & 8 \\
    VLEN8k-V40-S16 & 8192 & 40 & 16 \\
    VLEN4k-V48 & 4096 & 48 & 0 \\
    VLEN4k-V40-S16 & 4096 & 40 & 16 \\
    VLEN4k-V40-S8 & 4096 & 40 & 8 \\
    \hline
  \end{tabular}
\end{table}

\subsection{Physical Implementation and Power Estimation}\label{eval:synthesis}
We synthesize all the configurations in \Cref{table:configs} using Synopsys Design Compiler\textregistered, targeting the TSMC N4C 4-nanometer technology node \cite{tsmcn4c}. We enable the tool to automatically insert clock-gated cells to mitigate dynamic power consumption of all the design configurations. The \gls{vrf} of each configuration is generated as register file macros through the memory compiler portfolio provided as part of the technology \gls{pdk}.

After synthesis, the resulting gate-level netlist is used to perform \gls{gls} with back-annotated timing information. The \gls{gls} is driven by the verification environments described in \Cref{eval:setup}, which provide instruction streams representative of the evaluated workloads. During simulation, switching activity is captured and exported as activity files.

The outcome of the \gls{gls} is processed through Synopsys PrimePower® to estimate the total power consumption. Power estimates therefore account for realistic switching activity derived from the executed workloads and reflect the timing characteristics of the synthesized design.

Unless otherwise stated, reported power results correspond to the average total power of the \gls{vpu} during workload execution. Area results are obtained directly from the synthesis reports produced by Design Compiler.

\subsection{Workloads}\label{eval:benchmarks}
The evaluated \gls{vpu} targets application domains such as \gls{hpc} and AI/ML. To assess the impact of the proposed optimization across representative workloads, we evaluate the design using vectorized kernels derived from existing RISC-V vector benchmark suites. In particular, we consider applications derived from the RiVEC Benchmark Suite \cite{Ramirez2020}, which provides data-parallel workloads targeting \gls{rvv}, and from the software ecosystem described in published papers on vector architectures \cite{Cavalcante2020}. 

Performance evaluation is conducted through \gls{rtl} simulation of the \gls{vpu}, which provides cycle-accurate visibility into the microarchitectural behavior of the design. Given the computational cost of \gls{rtl} simulation for large hardware designs, the evaluation focuses on representative regions of interest extracted from the selected benchmark applications. The simulation environment operates in a bare-metal configuration without operating system support; therefore, benchmark kernels are adapted to run as standalone programs, removing dependencies on runtime libraries while preserving the core computational patterns of the original applications. Here we provide a brief description of each evaluation kernel:
\begin{itemize}
    \item \textit{gemm} implements a vectorized matrix multiplication in the form $C = \alpha AB + \beta C$, multipying a MxN matrix by a NxK matrix. We generate a version of the kernel with an unrolling factor of four for the innermost loop, matching the four \gls{fpus} available in each vector lane of the baseline design.
    \item \textit{dotproduct} computes element-wise vector multiplications followed by vector reductions.
    \item \textit{particle filter} implements a sequential algorithm to estimate the state of a dynamic system through particle propagation.
    \item \textit{layernorm} computes the layer normalization operation commonly used in deep learning models.
    \item \textit{lavaMD} calculates particle potential and relocation due to mutual forces between particles within a large 3D space.
    \item \textit{blackscholes} evaluates the closed-form Black-Scholes equation to price European-style financial options.
    \item \textit{swaptions} estimates the price of a portfolio of swaptions based on Monte Carlo methods.
    \item \textit{somier} simulates the dynamics of a deformable three-dimensional mass–spring system by computing spring forces and updating particle positions and velocities.
    \item \textit{axpy} implements a scaled vector addition as $Y = \alpha X + Y$.    
\end{itemize}

The kernels, listed with the evaluated input sizes in \Cref{table:kernels}, are compiled using the RISC-V GNU toolchain and manually inspected to ensure compatibility with the simulation environment. 
Together, these kernels represent common computation patterns found in HPC and AI/ML applications while exercising the instruction types targeted by the proposed optimization, including scalar-vector moves and vector reductions.

\begin{table}[htb]
  \centering
  \caption{Evaluation benchmarks.}
  \label{table:kernels}
  \begin{tabular}{lcc}
    \hline
    \textbf{Kernel} & \textbf{Application Domain} & \textbf{Input Size} \\
    \hline
    gemm & HPC, AI/ML & 256x256 matrices \\
    layernorm & HPC, AI/ML & 256x4096 elements\\
    dotproduct & HPC & 400000 elements \\
    axpy & HPC & 16384 elements \\
    lavaMD & Scientific Computing & 65536 atoms \\
    particle filter & Scientific Computing & 50000 particles \\
    blackscholes & Financial Computing & 4096 options \\
    swaptions & Financial Computing & 4096 paths \\
    \hline
  \end{tabular}
\end{table}

\subsection{Metrics}\label{eval:metrics}

The primary objective of this work is to improve the efficiency of the \gls{vpu} by redirecting the outputs of scalar-semantics vector instructions into the dedicated \gls{svrf}. To assess the effectiveness of this optimization, we focus on the following metrics:
\begin{enumerate}
    \item \textbf{Performance Impact.} To evaluate the performance impact of the SVRF, we compare the execution time of the proposed configurations against the baseline VPU across the considered benchmarks. The SVRF can reduce VRF pressure by storing scalar-semantics vector results in scalar physical registers, while rename-resolved operations can complete before reaching the issue stage. Consequently, the performance impact depends on the occurrence of scalar-semantics vector instructions and their position in the instruction stream. We evaluate these effects across the benchmark suite while varying the number of physical scalar and vector registers.
    \item \textbf{Area Savings.} Reducing vector register usage in favor of scalar register allocation can enable a smaller physical \gls{vrf} without compromising performance. Area is measured from synthesis reports of each configuration in \Cref{eval:configs}. This metric captures the potential silicon savings of the proposed design.
    \item \textbf{Power Consumption.} Because scalar-producing instructions no longer access the full-width vector registers, the dynamic power consumption in the \gls{vrf} is reduced. Moreover, configurations allowing for a smaller \gls{vrf} yield proportional reduction in static power consumption. Power is estimated using gate-level simulation as described in \Cref{eval:synthesis}. This metric provides insight into the total power benefits given by the \gls{svrf}.
\end{enumerate}

These metrics allow for a comprehensive assessment of the trade-offs and benefits of introducing the \gls{svrf} and the related register renaming mechanisms, quantifying its effect on resource utilization, silicon cost, power efficiency, and workload execution.

\begin{figure*}[htb] 
    \centering

    \begin{subfigure}{0.32\textwidth}
        \centering
        \includegraphics[width=\linewidth]{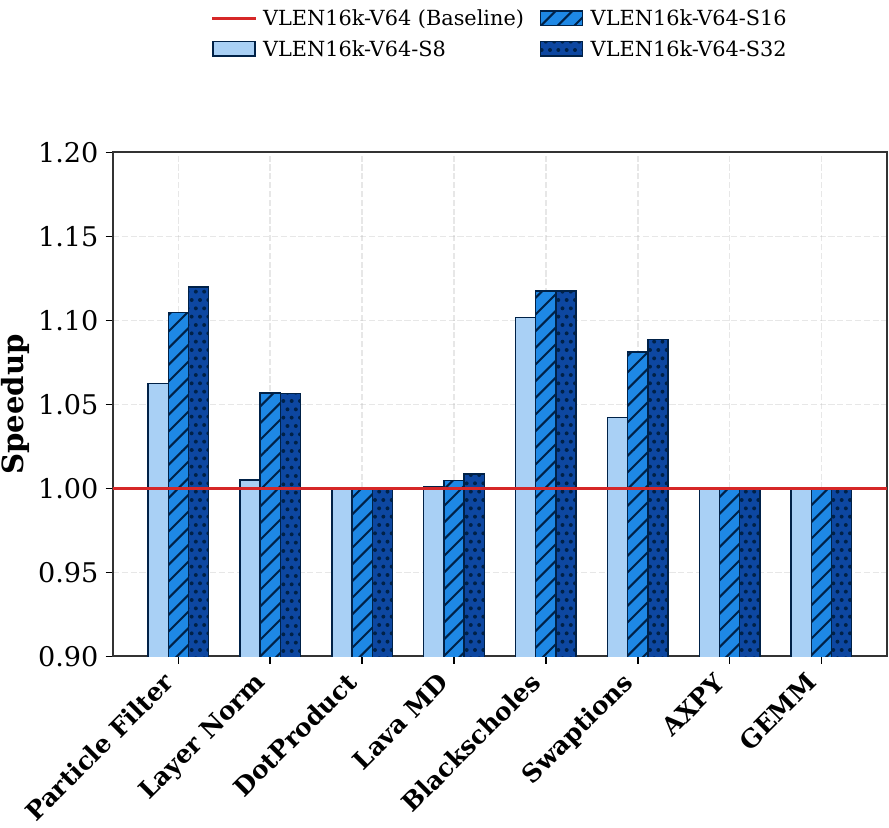}
        \caption{64 Vector Registers}
        \label{fig:speedup_64}
    \end{subfigure}
    \hfill 
    \begin{subfigure}{0.32\textwidth}
        \centering
        \includegraphics[width=\linewidth]{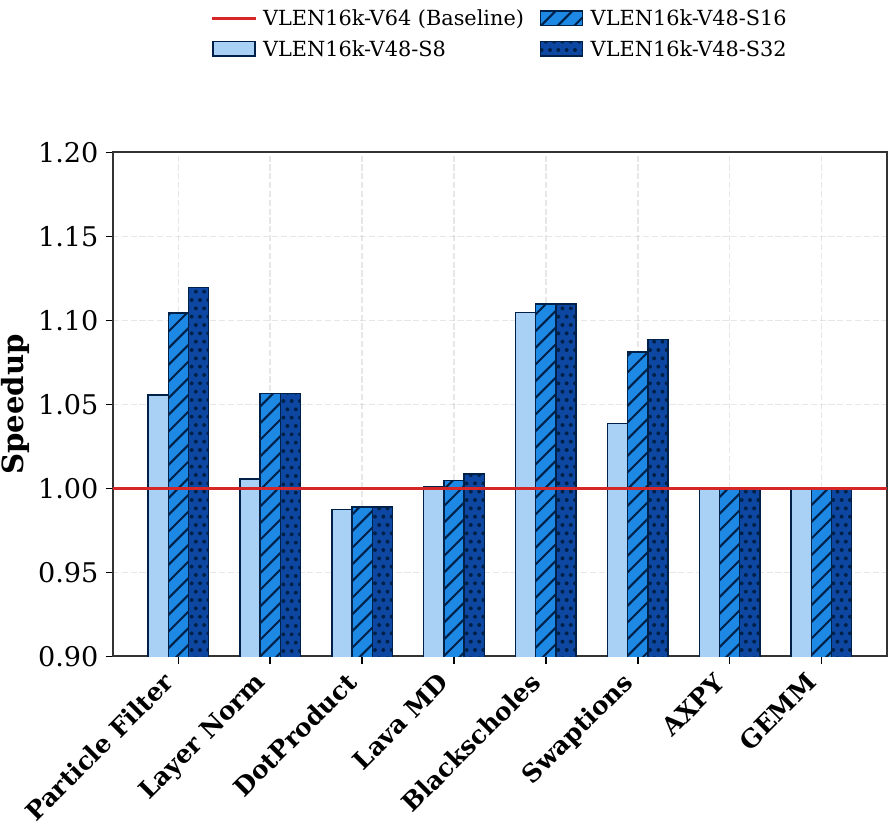}
        \caption{48 Vector Registers}
        \label{fig:speedup_48}
    \end{subfigure}
    \hfill
    \begin{subfigure}{0.32\textwidth}
        \centering
        \includegraphics[width=\linewidth]{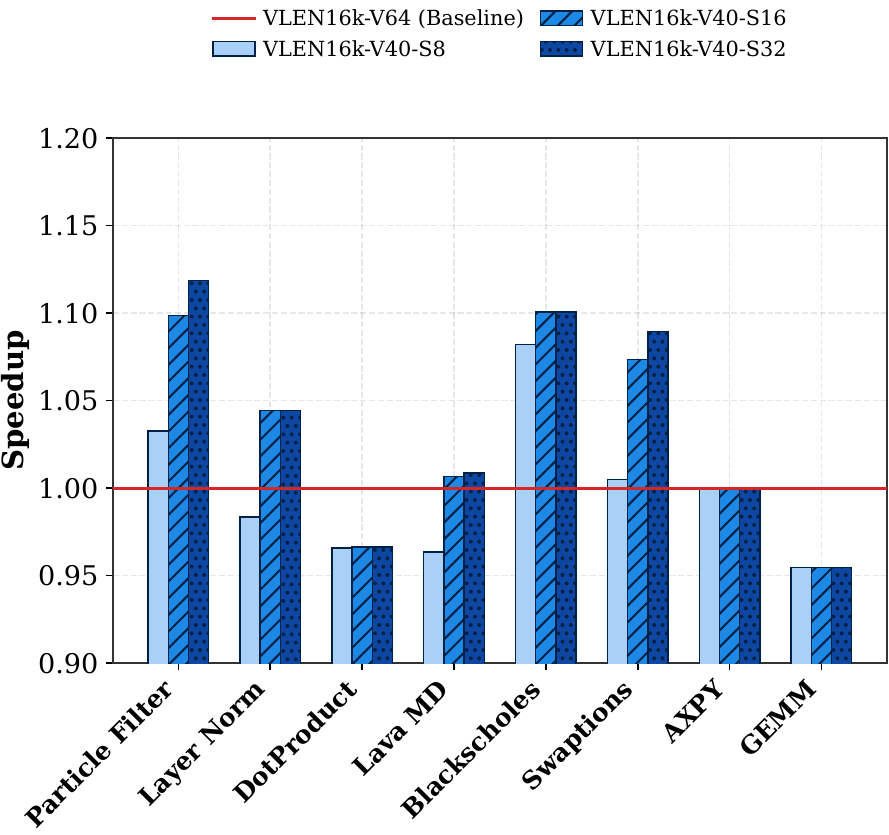}
        \caption{40 Vector Registers}
        \label{fig:speedup_40}
    \end{subfigure}

    \vspace{2mm} 
    \caption{Performance of the SVRF-enabled designs over the baseline configuration across different VRF sizes (64, 48, 40).}
    \label{fig:all_speedups_comparison}
\end{figure*}

\section{Evaluation}\label{evaluation}

This Section evaluates the effectiveness of the proposed scalar register file architecture in improving vector register utilization and reducing hardware cost while preserving performance. The evaluation focuses on the metrics defined in \Cref{eval:metrics}.

Experiments are conducted using the benchmarks described in \Cref{eval:benchmarks} across multiple hardware configurations. As detailed in \Cref{eval:configs}, these configurations vary the number of physical vector registers and scalar registers, as well as the hardware supported vector length. Unless otherwise specified, results are reported relative to the baseline \gls{vpu} configuration presented in \Cref{vitruvius}, which implements a conventional \gls{vrf} with no additional scalar register file support.





\subsection{Performance Impact}\label{results:performance}
The primary goal of the proposed \gls{svrf} is to improve the efficiency of vector register usage by avoiding the allocation of full vector registers for scalar-producing instructions. As a result, the mechanism is designed to preserve the execution performance of the baseline architecture while improving hardware efficiency. In this Section, we evaluate the execution performance of the optimized \gls{vpu} across the workloads described in \Cref{eval:benchmarks}.

\Cref{{fig:all_speedups_comparison}} reports the performance of the \gls{svrf}-enabled designs over the baseline configuration of the \gls{vpu}. Overall, the results show that the proposed mechanism preserves the performance of the baseline architecture across the evaluated workloads. As shown in \Cref{fig:speedup_64}, the majority of the benchmarks exhibit execution times comparable to the VLEN16k-V64 baseline configuration, confirming that redirecting scalar values to the \gls{svrf} does not introduce any performance penalty. Notably, the benchmarks \textit{particle filter}, \textit{blackscholes}, and \textit{swaptions} achieve modest performance improvements when the \gls{svrf} is enabled. These benchmarks contain a significant number of scalar-to-vector moves instructions, either used for bookkeeping operations or accumulator initialization. In these applications, the result of a scalar-semantics vector instruction is consumed by up to three subsequent vector instructions. Consequently, both \textit{particle filter} and \textit{blackscholes} achieve a speedup of 1.10X and 1.12X over VLEN16k-V64, respectively, even when the \gls{svrf} is configured with only 16 scalar registers (VLEN16k-V64-S16). The \textit{gemm} benchmark deserves special consideration. Although scalar-to-vector moves are used to initialize the accumulator registers of the matrix multiplication, they are executed only once at the beginning of every new iteration, resulting in a negligible impact on overall performance. Furthermore, the baseline \gls{vpu} architecture  achieves 62 FLOP/cycle for the evaluated \textit{gemm} implementation, corresponding to 97\% of its theoretical peak performance, leaving little opportunity for additional speedup.

Interestingly, configurations with fewer physical vector registers also benefit from the introduction of the \gls{svrf}, as shown in \Cref{fig:speedup_48,fig:speedup_40}. In particular, both VLEN16k-V48-S16 and VLEN-16k-V40-S16 outperform the VLEN16k-V64 baseline for \textit{particle filter}, \textit{blackscholes} and \textit{swaptions} when the \gls{svrf} is configured with 16 scalar registers.
These results indicate that improving vector register utilization enables more compact register file organizations while maintaining---or even improving---execution performance.


Overall, the results indicate that the proposed mechanism achieves its primary goal of improving hardware efficiency without compromising performance, while occasionally providing additional performance benefits in workloads with a high proportion of scalar-producing vector instructions.

\subsection{Hardware Cost and Area Impact}\label{results:area}

We evaluate the hardware cost implications of introducing the \gls{svrf} in the baseline \gls{vpu} architecture. Additionally, the results in \Cref{results:performance} show that the improved utilization of vector registers enabled by the \gls{svrf} allows the \gls{vrf} to be reduced without affecting execution performance.

\begin{figure}[htb!]
    \centering
    \includegraphics[width=\linewidth]{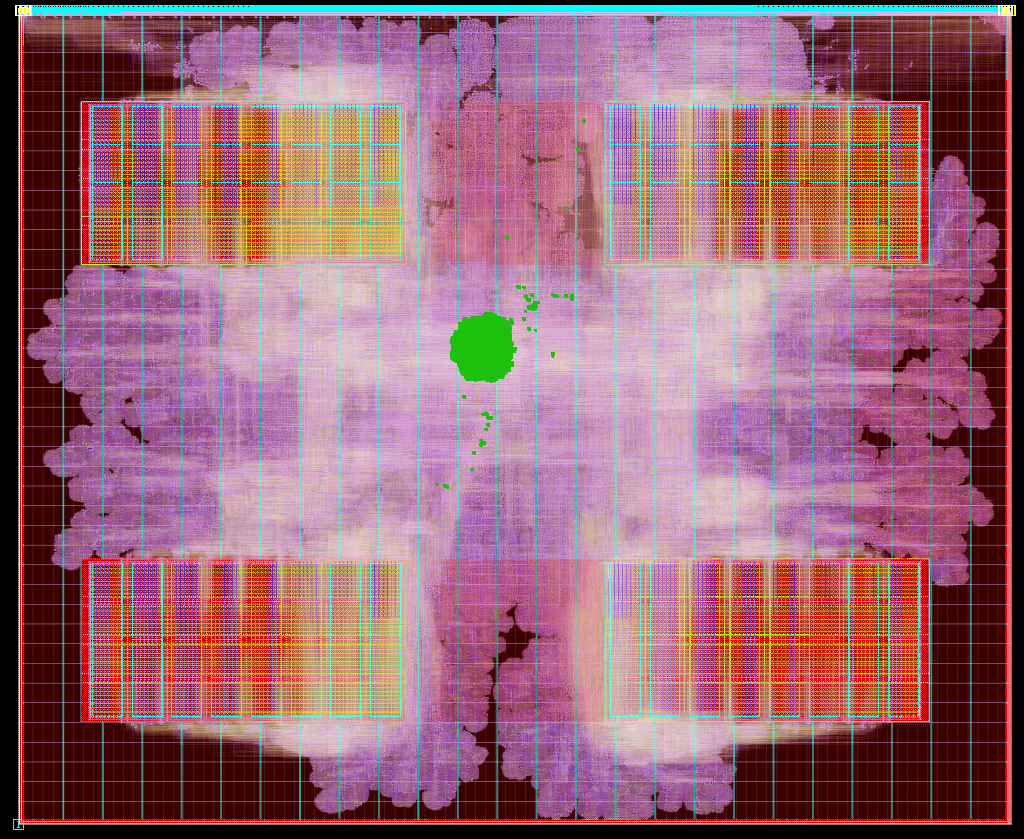}
    \caption{Floorplan of the VPU configured with 64 physical vector registers and 16 scalar registers in the SVRF. The SVRF is highlighted in green.}
    \label{fig:PnR}
\end{figure}

\begin{figure}[h!]
    \centering
    \includegraphics[width=\textwidth]{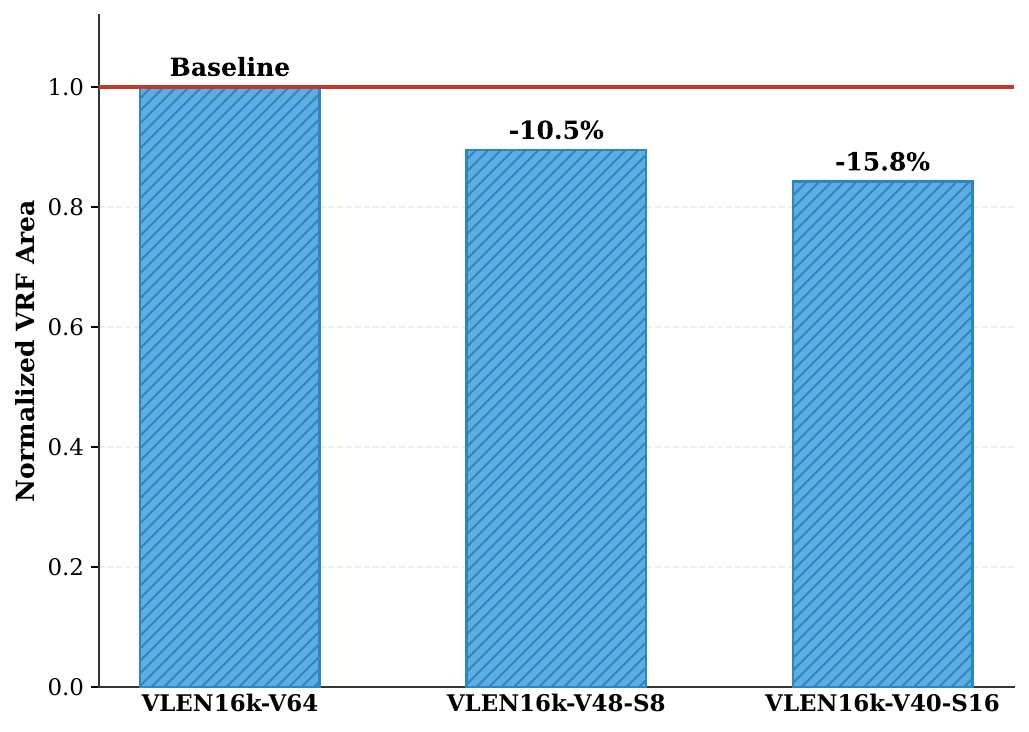}
    \caption{Normalized area of the VRF across the evaluated configurations. All values are normalized to the VLEN16k-V64 baseline.} 
    \label{fig:area_comparison}
\end{figure}

\begin{figure*}[htb] 
    \centering
    \includegraphics[width=\textwidth]{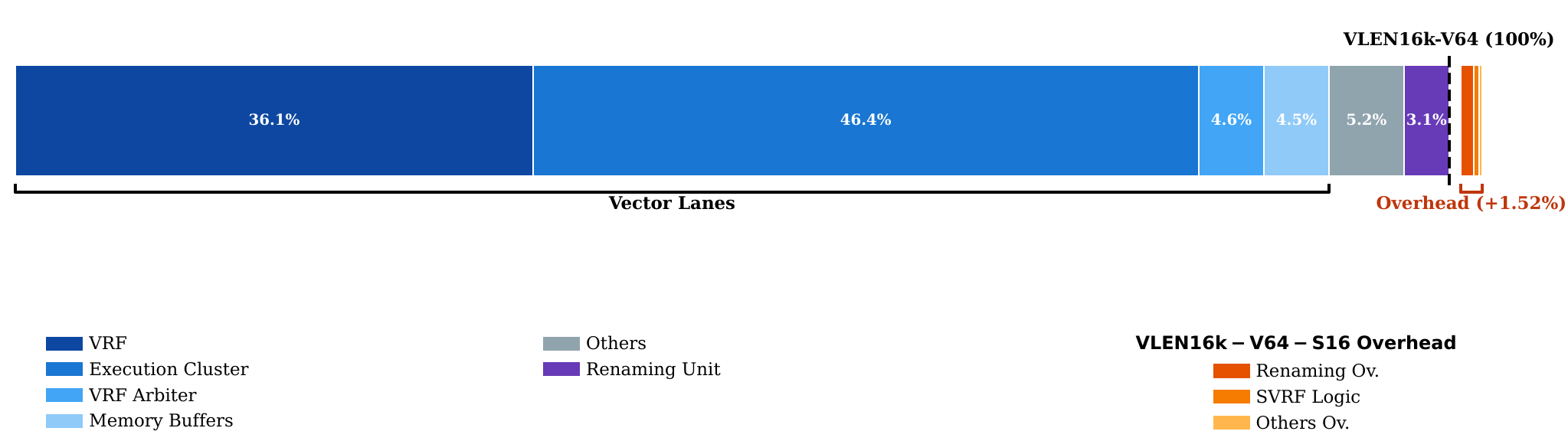}
    \caption{Area breakdown of the VPU architecture, including the overhead introduced by the VLEN16k-V64-S16 configuration relative to the baseline.}
    \label{fig:vpu_area}
\end{figure*}

\Cref{fig:PnR} shows the placed-and-routed design of the configuration VLEN16k-V64-S16 of the \gls{vpu}, highlighting the integration of the \gls{svrf} alongside the \gls{vrf}. In this configuration, the \gls{svrf} occupies only the 0.34\% of the total area. \Cref{fig:vpu_area} reports the area breakdown of the same configuration, splitting the contributions among the internal components of the \gls{vpu}. The modifications to the register renaming stage, together with the additional hardware required to integrate the \gls{svrf}, introduce a negligible area overhead of only 1.52\% with respect to the baseline architecture. As is typical of long-vector architectures, the \gls{vrf} is notably one of the dominant contributors to the total \gls{vpu} area.

\Cref{tab:area_savings} reports the post-synthesis area results for the evaluated configurations. Introducing the \gls{svrf} while preserving the original \gls{vrf} organization slightly increases the total area due to the additional scalar storage structure. However, when the \gls{svrf} is combined with a reduction in the number of physical vector registers, enabled by the observations presented in \Cref{results:performance}, the overall design becomes more area-efficient. In particular, the configuration VLEN16k-V40-S16 achieves the largest area reduction, decreasing the total area by up to 6\% compared with the baseline design. The resulting area savings can be reinvested in additional computational resources. For example, the recovered area is sufficient to integrate one additional FPU per vector lane, increasing the theoretical peak performance of the \gls{vpu} by 1.25X.

The benefits of integrating the \gls{svrf} become even more evident when considering the area savings enabled by \gls{vrf} reduced size. As shown in \Cref{fig:area_comparison}, the VLEN16k-V40-S16 configuration reduces the \gls{vrf} area by a 15.8\% while maintaining the same performance as the baseline design.

These results demonstrate that the improved utilization of vector registers enabled by the SVRF allows for a significantly more compact VRF organization without compromising execution performance.

\begin{table}[h!tb]
\centering
\caption{Total Area Analysis and Percentage Savings.}
\label{tab:area_savings}
\begin{tabular}{lcc}
\toprule
\textbf{Configuration} & \textbf{Total Area [$\mu m^2$]} & \textbf{Difference (\%)} \\
\midrule
VLEN16k-V64 (Baseline) & 427348.44 & --- \\
VLEN16k-V64-S32        & 437715.25 &  +2.43\\
VLEN16k-V64-S16        & 433795.49 &  +1.51\\
VLEN16k-V64-S16        & 432295.12 & +1.16 \\
VLEN16k-V48-S16        & 410403.02 & -3.97 \\
VLEN16k-V48-S8         & 410302.42 & -3.99 \\
VLEN16k-V40-S16        & 401548.82 & -6.04 \\

\bottomrule
\end{tabular}%
\end{table}

\begin{figure*}[htb]
    \centering
    \includegraphics[width=\textwidth]{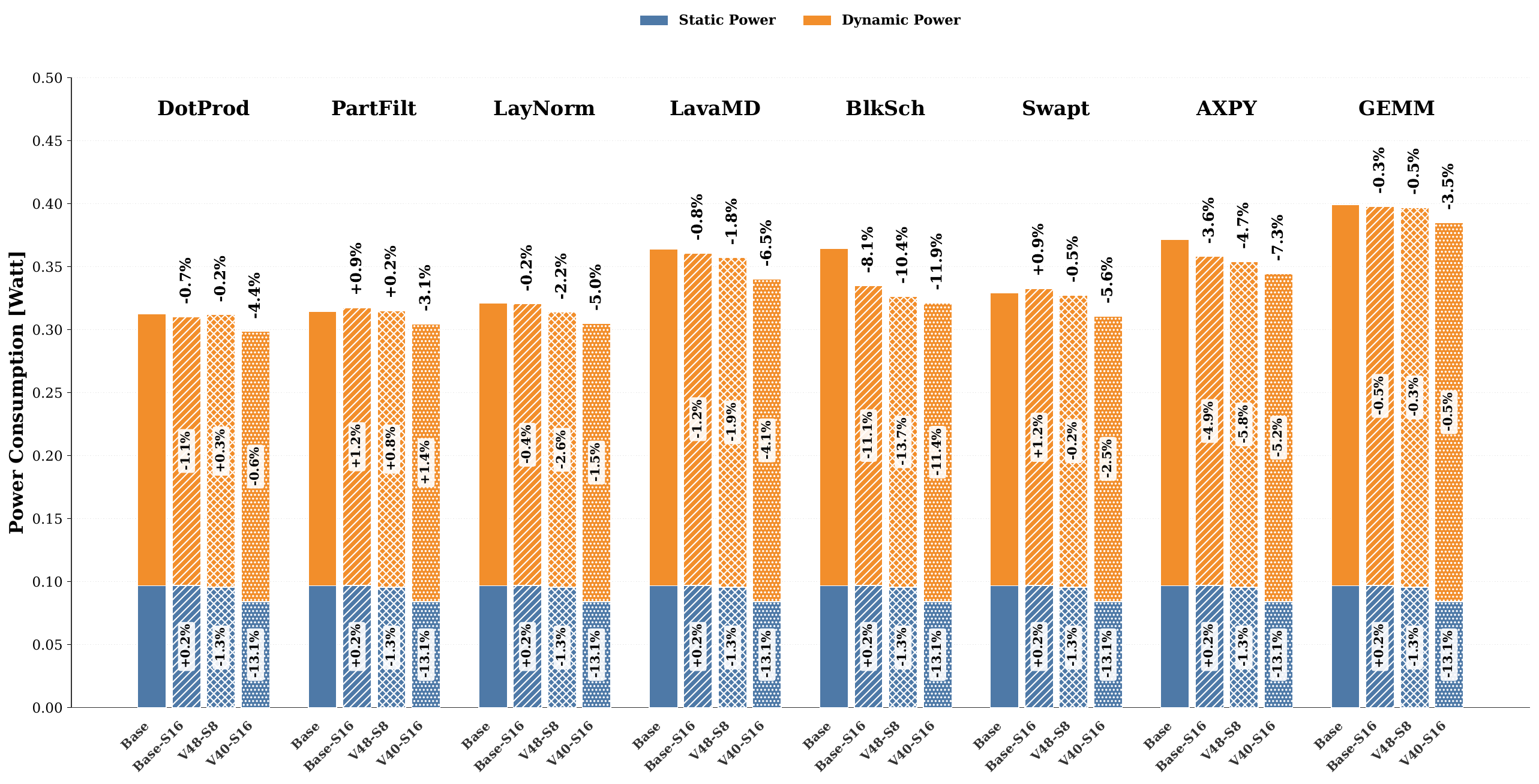}
    \caption{Power consumption breakdown for the evaluated workloads. The chart compares the baseline VLEN16k-V64 architecture with several optimized configurations featuring the SVRF (Scalar-Vector Register File). Each bar illustrates the contribution of static leakage and dynamic switching power in Watts.}
    \label{fig:full_power_analysis}
\end{figure*}

\subsection{Power consumption}

We evaluate the benefits of the \gls{svrf} architecture on the power consumption across all the evaluated workloads. Overall, the proposed \gls{svrf} architecture significantly reduce power consumption, as reported in \Cref{fig:full_power_analysis}. 

Dynamic power consumption is primarily reduced by eliminating the overhead of writing scalar results to full vector registers in the \gls{vrf}. In fact, rename-resolved instructions commit their results at the renaming stage, so they do not allocate entries in the instruction queues. Consequently, these instructions are not issued to the execution stage, and no \gls{vrf} write access occurs in the vector lanes. Additional dynamic power is saved when values previously stored in the \gls{svrf} are retrieved by subsequent instructions, as accessing data directly from the \gls{svrf} bypasses \gls{vrf} read operations. The best result is achieved on the \textit{blackscholes} benchmark, which reports a reduction of dynamic power consumption of 8.4\% with the configuration VLEN16k-V64-S16.

Configurations of the \gls{svrf} enabling a reduction of the \gls{vrf} size proportionally reduce static power consumption. For instance, static power consumption is reduced by up to 13\% for VLEN16k-V40-S16 and 1.3\% for VLEN16k-V48-S8, respectively, thanks to the area savings reported in \Cref{results:area}.





\subsection{Impact of Vector Length}
 While the baseline \gls{vpu} targets long-vector execution (VLEN=16384), it is also important to evaluate whether the proposed \gls{svrf} remains effective for shorter vector lengths. The results reported in this Section correspond to the VLEN8k and VLEN4k configurations described in \Cref{eval:configs}.

The exploration is limited to VLEN$\geq$4096. Reducing VLEN further would require more substantial modifications to the microarchitecture, such as restructuring the \gls{vrf} or altering the lane configuration. As a result, lower values of VLEN fall outside the design space of the current implementation and are left for future investigation.

\begin{figure}[h!tb]
     \centering
     \begin{subfigure}{0.8\columnwidth}
         \centering
         \includegraphics[width=\linewidth]{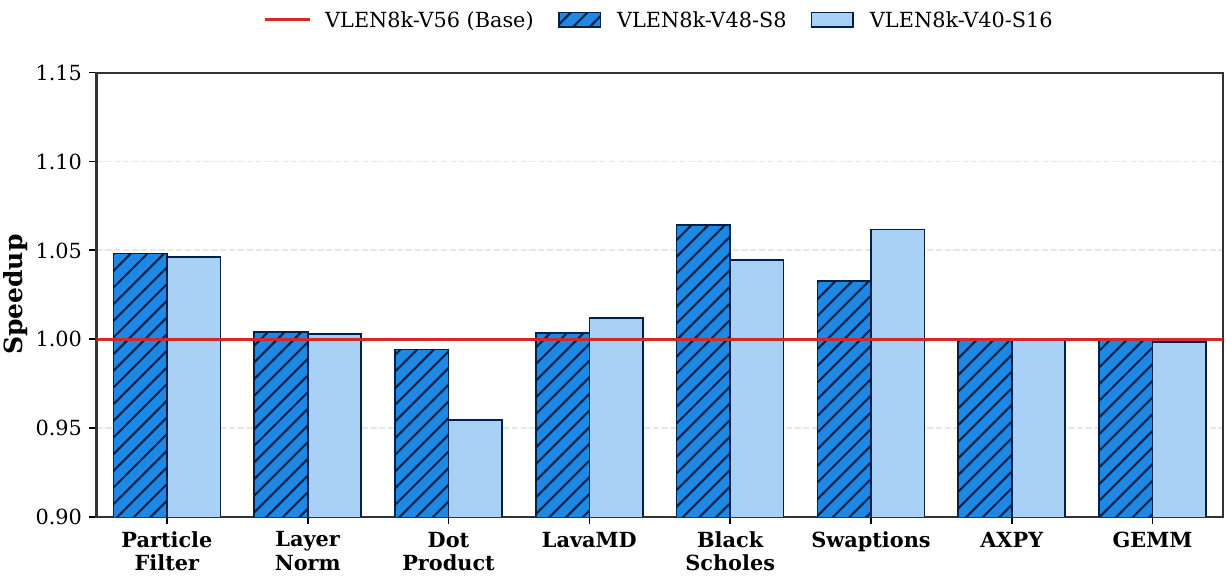}
         \caption{Analysis for VLEN 8k configurations.}
         \label{fig:vlen8k}
     \end{subfigure}
     
     \vspace{0.5cm} 
     
     \begin{subfigure}{0.8\columnwidth}
         \centering
         \includegraphics[width=\linewidth]{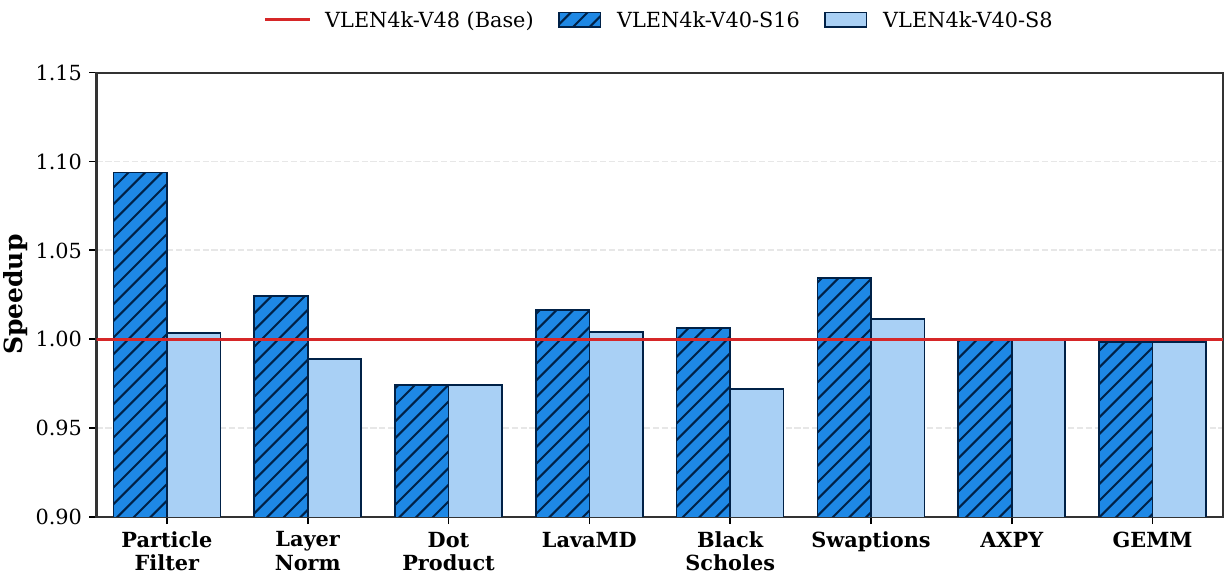}
         \caption{Analysis for VLEN 4k configurations.}
         \label{fig:vlen4k}
     \end{subfigure}
\caption{Normalized performance comparison of area-reduced VRF configurations relative to the baseline for VLEN=8k (a) and VLEN=4k (b).}
     \label{fig:performance_analysis}
\end{figure}

\Cref{fig:performance_analysis} compares the performance of the smaller-VLEN baselines with the corresponding configurations incorporating the reduced \gls{vrf} and the integrated \gls{svrf}. The results show that the mechanism continues to provide benefits for shorter vector lengths, albeit not as pronounced as for VLEN=16384. The integration of the \gls{svrf} still enables a reduction in the number of physical vector registers for both VLEN=8192 and VLEN=4096 while maintaining comparable execution performance.

\begin{figure}[htb] 
    \centering
    \includegraphics[width=\textwidth]{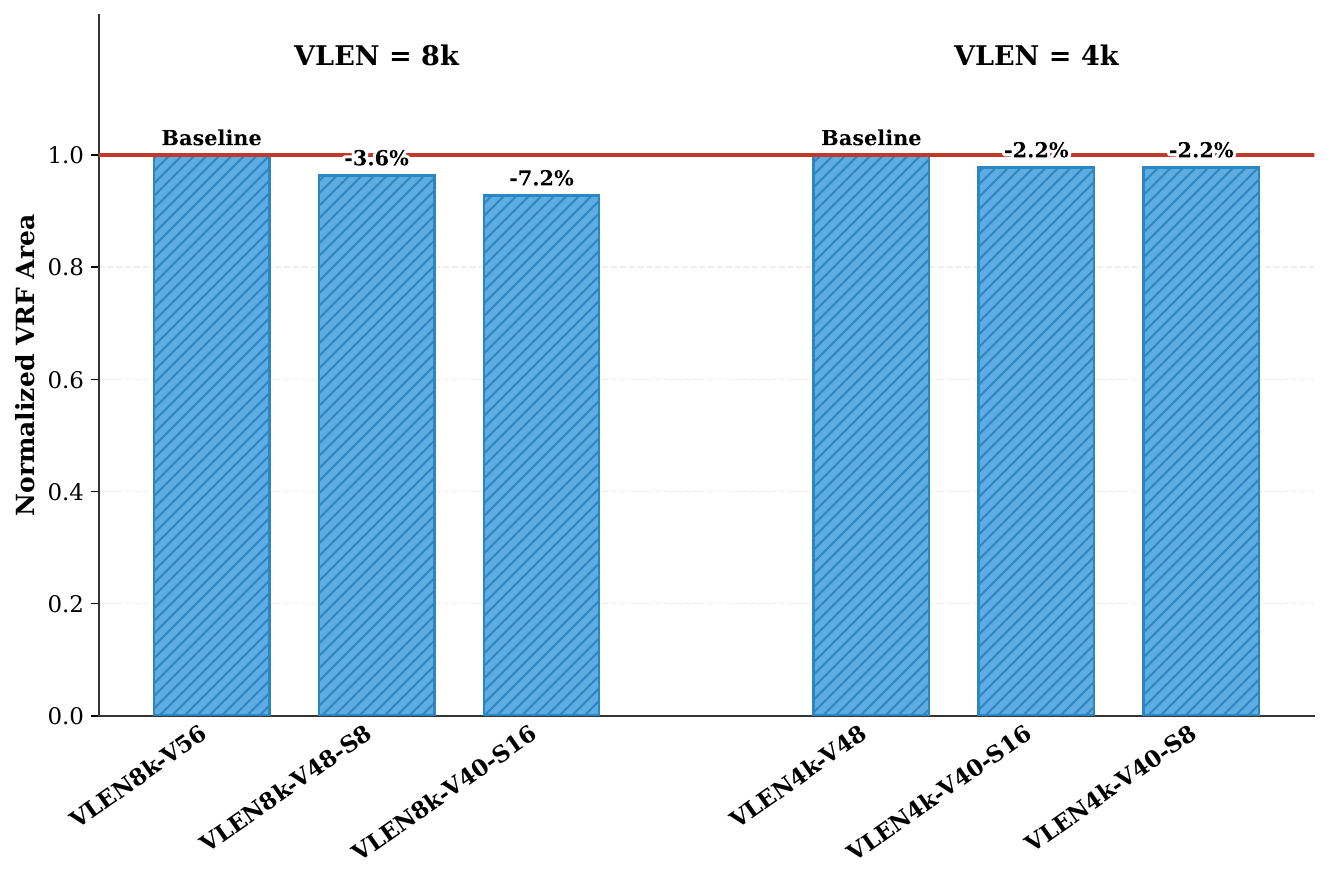}
    \caption{Normalized VRF area comparison for VLEN=8k and VLEN=4k configurations. Area savings are calculated relative to the respective baselines.}
    \label{fig:vrf_area_savings}
\end{figure}

The area results for smaller-VLEN configurations are reported in \Cref{fig:vrf_area_savings}. As expected, the area savings are less pronounced than for VLEN=16384. The largest reduction is achieved for the configuration VLEN8k-V40-S16, where decreasing the number of physical registers from 56 to 48 reduces the \gls{vrf} area by up to 7.2\%. The area savings are smaller for VLEN=4096 because each physical vector register contribute less to the overall design area as vector length decreases. Consequently, the \gls{vrf} accounts for a smaller fraction of the total \gls{vpu} area, limiting the impact of further reductions in the number of physical registers.

\section{Related Work}\label{sota}

Designing a performant and efficient register file has long been recognized as a critical challenge in computer architecture \cite{Zyuban1998}\cite{Balasub2001}. This is particularly true in vector processors due to its impact on area, power consumption, and scalability \cite{Arima2021}\cite{Kozyrakis2003}\cite{Cavalcante2022}. As vector lengths increase, each physical vector register becomes wider and more expensive to access, making the \gls{vrf} one of the most costly structures in the design. 
Consequently, several works have explored techniques to improve the efficiency of large VRFs.

In AVA \cite{Lazo2022}, the authors propose a two-level register file organization which redistributes the vector registers according to the vector length requested by the target applications. Short vectors are mapped in the \gls{vrf}, while longer vectors are mapped across the \gls{vrf} and a reserved area in the memory hierarchy, termed "memory \gls{vrf}". A dedicated swap mechanism is used to bring in the \gls{vrf} only the vector registers that are needed by the executing instructions. Similarly, Titopoulos et al.\cite{Titopoulos2025} introduce a compact \gls{vrf} that dynamically caches only the most recently accessed vector registers, while the remaining registers are "dispersed" in the memory subsystem.
Other research has explored architectural techniques to reduce the frequency of accesses to large register files. The MANIC architecture \cite{Gobieski2019} proposes a vector-dataflow execution model that forwards intermediate values directly between functional units, reducing the reliance on a centralized register file. While these works focus on improving the storage efficiency and organization of the \gls{vrf}, they continue to treat vector results as full-width values that must reside entirely within the \gls{vrf}.

Challenges associated with the design of the \gls{vrf} are central in open-source implementations of \gls{rvv}-compliant vector processors. Ara2 \cite{Perotti2024} presents an efficient vector processor architecture that distributes the \gls{vrf} across vector lanes to sustain high memory bandwidth and computational throughput. Building on this design, AraXL \cite{Purayil2025} explores physically scalable vector processor implementations capable of supporting ultra-wide vectors while maintaining efficient execution. These processors focus on scalable datapath organization and efficient \gls{vrf} implementation for long-vector execution. However, as in conventional vector processors, all vector instruction results are stored in full-width vector registers regardless of their logical data footprint.

In contrast to these approaches, this work focuses on a different and largely unexplored inefficiency arising from scalar-semantics vector instructions. Operations such as scalar-to-vector broadcasts, scalar-indexed gathers, and vector reductions logically operate on scalar values or produce scalar results, yet conventional implementations still allocate full-width vector registers to represent these values. As vector lengths increase, this mismatch leads to unnecessary register pressure, increased \gls{vrf} activity, and higher energy consumption. To the best of our knowledge, prior work has not explored separating scalar-valued vector results from the \gls{vrf} itself. This work addresses this limitation by introducing the \gls{svrf}, a microarchitectural extension that decouples scalar-valued vector results from the full-width \gls{vrf}. The proposed mechanism is transparent to the instruction set architecture and can be applied to vector processors supporting register renaming independently of the specific vector \gls{isa}.

\section{Conclusion}\label{conclusion}
This paper presented the \gls{svrf}, a microarchitectural extension that improves the efficiency of vector processors by avoiding the allocation of full-width vector registers for scalar-valued vector results. 
The proposed \gls{svrf} stores these values in a dedicated scalar structure and distributes them to the vector lanes on demand, reducing unnecessary \gls{vrf} usage while preserving the semantics of the original vector instructions. The mechanism is entirely microarchitectural and can be integrated into vector processors equipped with vector register renaming, independently of the supported vector \gls{isa}. Future work includes exploring additional scalarization opportunities and evaluating the proposed mechanism in vector processors with different vector lengths, including embedded designs with stricter area and energy constraints.



\bibliographystyle{unsrt}  
\bibliography{refs}  

\end{document}